\documentclass[final,5p,twocolumn,authoryear]{elsarticle}
\usepackage[T1]{fontenc}
\usepackage[english]{babel}
\usepackage{natbib}
\usepackage[table]{xcolor}
\usepackage{float} 
\usepackage{amssymb}
\usepackage{amsmath}
\usepackage{lineno}
\usepackage{hyperref}          
\hypersetup{hidelinks}
\usepackage{ragged2e}
\usepackage[most]{tcolorbox}

\journal{Neuroscience and Biobehavioural Reviews}

\begin{document}
\sloppy
\begin{frontmatter}




\title{Evidence of Physiological Co-Modulation During Human-Animal Interaction: A Systematic Review} 


\author{Ginevra Bargigli\corref{c1}\fnref{f1}}
\ead{ginevra.bargigli@unifi.it}
\fntext[f1]{ComPBioS, Department of Information Engineering, University of Florence, Florence, Italy}

\author{Lorenzo Frassineti\fnref{f1}}
\ead{lorenzo.frassineti@unifi.it}

\author{Antonio Lanata\fnref{f1}}
\ead{antonio.lanata@unifi.it}

\author{Paolo Baragli\fnref{f2}}
\ead{paolo.baragli@unipi.it}
\fntext[f2]{Department of Veterinary Sciences, University of Pisa, Pisa, Italy.}

\author{Chiara Scopa\fnref{f2}}
\ead{chiara.scopa@vet.unipi.it}

\author{Aglaia Vignoli\fnref{f3}}
\ead{aglaia.vignoli@unimi.it}
\fntext[f3]{Department of Health Sciences, University of Milan, Milan, Italy}

\cortext[c1]{Corresponding author}

\affiliation{
    organization={University of Florence},
    addressline={P.za di San Marco, 4}, 
    city={Florence},
    postcode={50121}, 
    state={FI},
    country={Italy}
}

\affiliation{
    organization={University of Pisa},
    addressline={Viale delle Piagge 2},
    city={Pisa},
    postcode={56124}, 
    state={PI},
    country={Italy}
}

\affiliation{
    organization={University of Milan},
    addressline={Via Festa del Perdono, 7},
    city={Milan},
    postcode={20122}, 
    state={MI},
    country={Italy}
}

\begin{abstract}
This review examines the evidence in the literature for \textit{physiological co-modulation during human-animal interaction}. The aim of this work is to identify studies that assessed co-modulation via simultaneous measurement of physiological signals in both species, performing quantitative comparisons, and evaluate the consistency of the findings.\\
We searched PubMed, EMBASE, Scopus, Google Scholar, Animal Studies Repository, and the "Consensus app" tool (\url{https://consensus.app/}) between June and August 2025 (last search: August 5, 2025).\\
Risk of bias was assessed using an adapted version of the ROBINS-I V2 tool \citep{sterneROBINSIToolAssessing2016}.\\
Results were grouped by data analysis method, interaction context, and physiological parameter. The data were synthesised narratively, in structured tables and in barplots. Thirty-seven studies were included, primarily focusing on dogs (n = 22) and horses (n = 15), framed primarily within the interaction contexts of Animal-Assisted Therapy/Intervention (AAT/AAI) and companionship. Cardiac and hormonal measures were most frequently assessed. Most studies (n = 20) performed correlation analyses.
Sample sizes ranged from $\leq$ 10 to $\geq$ 130 dyads.
Co-modulation resulted significant in 22 studies, partial (limited to subsets of data) in 9, and absent in 6. Time-series coupling methods yielded more consistent evidence than discrete-time correlations.
Many studies had small samples and did not explicitly test for significant co-modulation.
Evidence, while not conclusive, supports physiological co-modulation during human-animal interactions. However, studies' heterogeneity limits generalizability: rather than indicating a universal phenomenon, findings suggest co-modulation may emerge under specific biological and methodological conditions. Future research should explicitly test its presence across contexts.\\
\end{abstract}


\end{frontmatter}






\section{Introduction}
\label{Introduction}
\subsection{Rationale}
\label{Rationale}
In the context of this review, \textit{"physiological co-modulation"} refers to the dynamic alignment of physiological signals between individuals during interaction \citep{koskelaBehavioralEmotionalComodulation2024k}. In human-animal interaction (HAI) research, this phenomenon has been increasingly investigated as a potential marker of emotional attunement, mutual regulation, and relational bonding, echoing human studies \citep{gollandMereCoPresenceSynchronization2015, kinreichBraintoBrainSynchronyNaturalistic2017, bizzegoStrangersFriendsLovers2019,  gordonPhysiologicalBehavioralSynchrony2020, linEmotionContagionPhysiological2024}. However, the conceptual and methodological landscape remains fragmented. Studies often adopt the term “synchrony” broadly, applying it to both time-resolved coupling and static associations, and relying on heterogeneous analytical approaches. Recent work has thus proposed the term co-modulation to describe mutual physiological influence over extended timescales, without implying strict temporal synchrony \citep{koskelaBehavioralEmotionalComodulation2024k}. This broader conceptualization accommodates the diversity of physiological signals and interaction contexts observed in HAI studies, ranging from therapeutic interventions and companionship to sport and working environments. Despite growing interest, the question of whether physiological co-modulation is a generalizable feature of human-animal interaction or a context-dependent phenomenon shaped by species, interaction context, sampling modality, and analytical method, remains unresolved.\\
\subsection{Objectives}
\label{Objectives}
To address this gap, we conducted a systematic review focused on studies that simultaneously measured physiological signals in humans and animals during interaction, and performed quantitative comparisons. By synthesizing existing evidence, our aim was to map the presence and methodological characteristics of physiological co-modulation across diverse interaction contexts, species, and analytical frameworks. The overall goal is to clarify the conceptual field and inform future research on the dynamics of interspecies physiological co-modulation.\\
The aim of this study is coherently represented through the PICO framework (\hyperref[Box1]{Box 1}), which provides a structured approach to delineating the population, intervention, comparison, and outcomes of a systematic review \citep{eriksenImpactPatientIntervention2018}.\\

\begin{tcolorbox}[colframe=black, title= Box 1: PICO framework for this review]
\RaggedRight
\footnotesize
\label{Box1}
\textbf{Population:} Humans and animals engaged in interaction.

\textbf{Intervention/Exposure:} Simultaneous measurement of physiological parameters during human-animal interaction.

\textbf{Comparator:} No interaction, asynchronous measurement, or baseline.

\textbf{Outcome:} Reported evidence of physiological co-modulation.
\end{tcolorbox}
\normalsize

\section{Methods}
\label{Methods}
\subsection{Eligibility Criteria}
\label{eligibility criteria}
Eligibility criteria were defined based on the review's research question, as well as the PICO framework of this work (\hyperref[Box1]{Box 1}).\\
Neither the human nor the animal populations to be included in the studies were subjected to any restrictive criteria: studies involving any human participants (regardless of age, sex, or health status) and any animal species were considered eligible.\\ 
Only studies that simultaneously measured physiological parameters in both humans and animals during an interaction were considered eligible. Simultaneity was broadly interpreted as physiological data being recorded in overlapping time windows, with a temporal resolution sufficiently similar to allow a comparison. No minimum or maximum duration of physiological recording was required for inclusion.\\
No restrictions were applied regarding study design; both observational and experimental studies were considered eligible. Moreover, the type of interaction, as well as the setting, were not subject to any restriction criteria.\\
No restrictions were applied to the specific physiological parameters measured or the technical features of data acquisition. However, the search strategy included terms related to commonly used physiological measures (e.g., electroencephalography (EEG), photoplethysmography (PPG), functional Near-Infrared Spectroscopy (fNIRS), heart rate, oxytocin, cortisol, breath) to enhance sensitivity.\\
No formal comparator was required for inclusion. Notably, comparators were not consistently present across included studies. When present, although classifiable in general as "no interaction" conditions, comparators were heterogeneous and highly specific to individual study designs, further limiting their utility for synthesis purposes.\\
Studies were included if they performed any \textit{quantitative analysis to assess co-modulation}, interpreted as \textit{any comparative statistical evaluation between variations of human and animal physiological signals}. Thus, papers that measured physiological parameters in both humans and animals but did not quantitatively compare those measures, performing separate analysis pipelines, were excluded. Co-modulation was broadly defined to include both time-resolved coupling and static associations.\\
To ensure methodological rigour and data availability, papers that had not been peer reviewed, master's or bachelor's theses, unpublished papers, conference papers and papers written in languages other than English were excluded.\\
In performing citation searching a filter for papers published after 2024 was applied in order to limit the inclusion to the most recent advances on the subject. No other restrictions were applied regarding the year of publication.\\
Studies in which the outcome of interest was not measured at all were excluded in the title-screening step of the selection; exclusions attributed to the absence of outcome explicit report were performed in the text-screening step of the selection.\\
No formal protocol was registered for this review. However, all methodological decisions, including post hoc amendments, were transparently reported in accordance with PRISMA 2020 guidelines \citep{pagePRISMA2020Explanation2021}.\\

For the synthesis, data were grouped according to three main dimensions: the data analysis method (e.g., time-series coupling analysis, regression analysis, correlation analysis), the context of the interaction (e.g., Animal Assisted Therapy/Intervention, Companionship, Agonistic Sport, Non-agonistic Sport, Working Animals), and the measured physiological parameter (e.g., Cardiac Activity, Hormones, Others). A further subdivision was made based on the animal species enrolled in the studies (e.g. Dog, Horse, Others).\\
These dimensions were adopted to adequately represent the heterogeneity observed across studies, particularly along three dimensions of the PICO framework (\hyperref[Box1]{Box 1}): Population (animal species), Intervention (context of interaction and physiological parameter measured), and Outcome (data analysis method).\\ 
For each of these groups and their respective subgroups, the synthesis presents the distribution of co-modulation outcomes --- classified into three categories: significant, partial (e.g., limited to specific conditions or subgroups), or absent, based on the authors' interpretation of their statistical analyses --- across one or more of the other dimensions, and the corresponding risk of bias assessments.\\

\subsection{Information Sources}
\label{information sources}
The following databases were searched: PubMed, EMBASE, Scopus, Google Scholar, Animal Studies Repository.
Additionally, the research tool "Consensus app" (\url{https://consensus.app/}) was interrogated. Although not a traditional bibliographic database, the Consensus App was included to enhance sensitivity and identify potentially relevant studies not indexed in standard databases \citep{apataUseGenerativeArtificial2025}. Its use was exploratory and the top-ranked results were screened with the same eligibility criteria that were adopted for the databases.\\
Moreover, citing search was performed for each eligible entry identified through database and Consensus App tool interrogation (via Google Scholar web interface). A filter for papers published after 2024 was applied in order to limit the inclusion to the most recent advances on the subject. The 'related articles' function in the Google Scholar web interface was analogously used.\\
In order to identify additional relevant studies, faculty members, doctoral and postdoctoral researchers, and research fellows affiliated with the Computational Physiology and Biomedical Systems (CoMPBioS) research group from the Department of Information Engineering and related units at the University of Florence, as well as external collaborators and affiliated researchers working with CoMPBioS on related projects, were contacted and invited to share any publications they were aware of that met the inclusion criteria of this review.
All details on the information sources are reported in the PRISMA search strategy document of this review (see: Supplementary Document S1).

\subsection{Search Strategy}
\label{search strategy}
All searches were conducted manually by entering each Boolean keyword combination directly into the search bar of each database's web interface, without the use of advanced syntax, field tags, or controlled vocabulary. As an example, the following query was entered directly into the search bar of each database:\\
\texttt{"human animal interaction" AND "physiological measures"}.\\
Separate searches were conducted for each keyword combination to facilitate workload distribution, maintain syntactic clarity, and ensure reproducibility. This approach also allowed for better control over search results, as some databases handle complex Boolean logic differently, and minimized the risk of retrieving irrelevant records due to overly broad Boolean logic.
The Consensus App was interrogated with prompts written in natural language and structured as questions, with which it best performs.\\
The initial keywords combination ("human animal interaction" AND "physiological measures") and Consensus App prompt ("are there publications which correlate animal and human physiological measures measures in the context of human-animal interaction?") were derived from the Intervention feature of the PICO (\hyperref[Box1]{Box 1}) of this review (Simultaneous measurement of physiological parameters during human-animal interaction).\\
As defined by \citep{frizeMeasuringPhysiologicalVariables2014}, the term "physiological" was interpreted as "\textit{physical quantities such as temperature, cardiac activity, blood pressure, chemistry values in the blood and urine, enzymes, and proteins}".\\
The choice of the composite keyword "human animal interaction" was based on its extensive use in the context of Animal Assisted Therapy (AAT) and Intervention (AAI)  literature.\\
The term “correlate” was used in a broad sense to capture studies that simultaneously measured physiological signals in humans and animals during interaction and performed any quantitative analysis assessing their co-modulation.\\
The decision to restrict the keyword strategy to the Intervention component of the PICO was aimed at maximizing the coverage of the search, in order to capture a wider spectrum of potentially relevant studies.\\
Then, in order to enhance the sensitivity of the search, additional queries were performed using more specific alternatives to the general term “physiological parameters”, namely the most common physiological signals in current human-animal interaction research (EEG, PPG, fNIRS, heart rate, oxytocin, cortisol, breath).\\
All detailed information on the search strategy is reported in the PRISMA Search strategy document of this review (see: Supplementary Document S1).

\subsection{Selection Process}
\label{selection process}
\begin{figure*}[htbp]
    \centering
    \includegraphics[width=\textwidth]{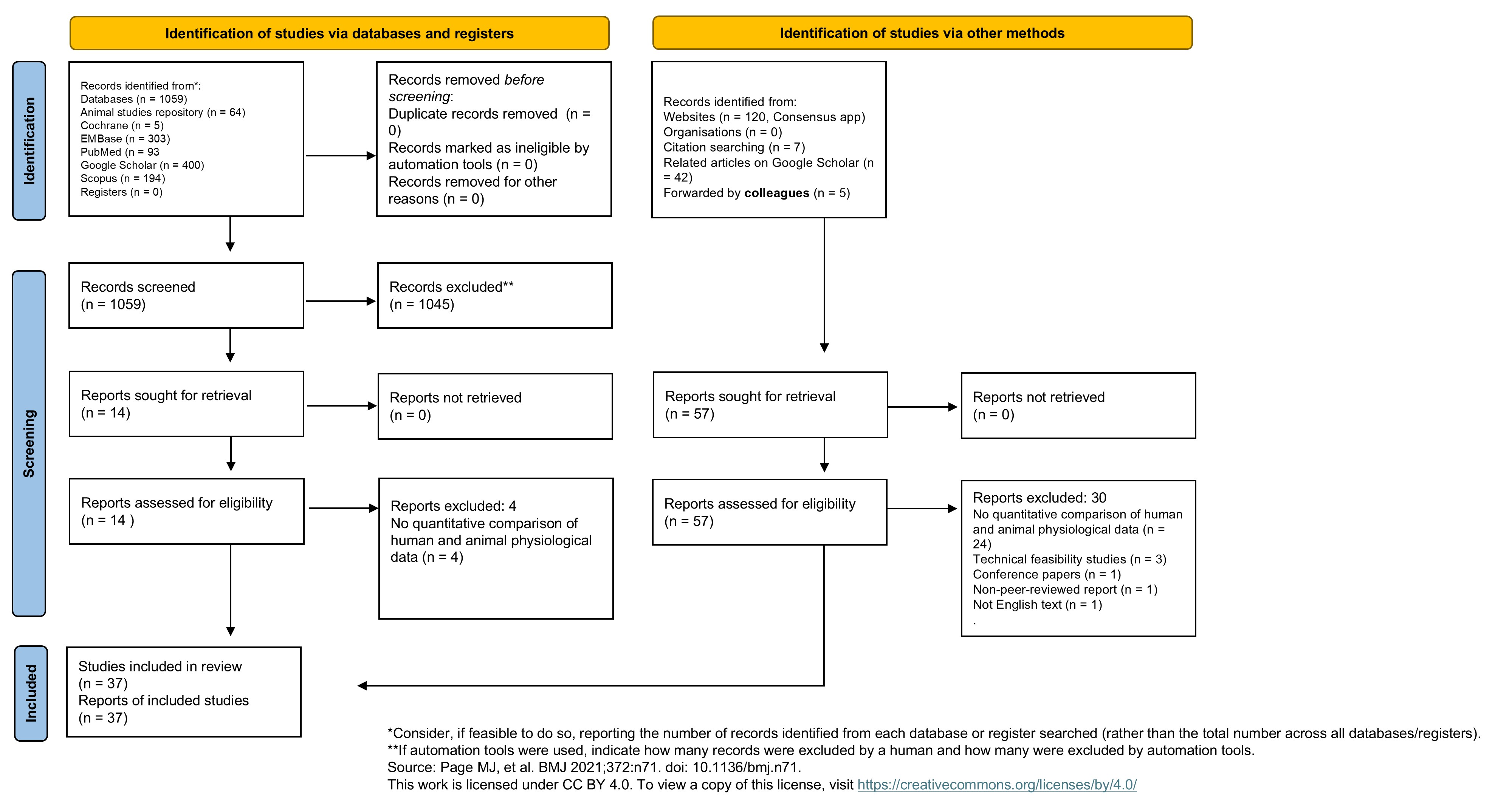}
    \caption{\textbf{PRISMA 2020 flow diagram} --- the diagram illustrates the number of records identified, screened, and included in this review.}
    \label{Fig1}
\end{figure*}

Two reviewers worked independently, performing separate searches. The second reviewer performed the search with the same procedure as the first reviewer. The order of magnitude for the results was analogous to the first.\\
Only one additional record \citep{malinowskiEffectsEquineAssisted2018b}, that satisfied the Intervention-based initial criterion of "reporting physiological measurements collected simultaneously in both humans and animals during interaction" was retrieved by the second reviewer. After text screening, performed by the first reviewer, the record was excluded as it did not fulfil the criterion of performing a co-modulation comparative statistical analysis.\\
No automation tools were used to screen entries.
Titles for a total of 1179 records were screened (1059 from bibliographic databases and 120 from the Consensus App). Titles that clearly referred exclusively to behavioural measures, or to physiological parameters measured in either humans or animals only, were not included. Titles that did not explicitly mention physiological measures in both species, but also did not clearly indicate exclusively behavioural or single-species data, were further assessed at the abstract level.\\ Only studies that reported physiological measurements collected simultaneously in both humans and animals during interaction were included at this level.\\
A total of 22 entries (14 from databases, 3 from Consensus, 5 shared by the individuals who had been invited to contribute) were initially included.\\ 
Citing search was performed for each eligible entry identified as well as interrogation of the 'related articles' tools in Google Scholar.\\
Additional 49 entries (7 from citing search, 42 from 'related articles') that reported physiological measurements collected simultaneously in both humans and animals were included.\\ The full-text of a total of 71 entries was thus screened.\\
Search results were saved as a Zotero collection (see: Supplementary File S2). The PRISMA Flow Diagram for this review (Figure \ref{Fig1}) provides a complete overview of the process.

\subsection{Data Collection Process}
\label{data collection process}
Data were extracted from the full texts of the included studies using a structured Excel spreadsheet, which served as the data collection form. The data collection form, including the extracted data, and the accompanying data dictionary are available as supplementary material (see: Supplementary File S3, Supplementary Document S4).\\
The first reviewer manually performed the data extraction. All extracted items were subsequently checked by the second reviewer. Discrepancies or uncertainties were resolved through iterative reading and cross-checking of the source documents. No translations were required. No data were extracted from figures using software tools. Information regarding the presence of behavioural and psychological assessments was extracted, but was not included in the analysis, as the focus of the review was the quantitative comparison of human and animal physiological parameters.\\

\subsection{Data Items}
\label{data items - outcomes}
\subsubsection*{Outcomes}
The outcome domain was defined as presence of physiological co-modulation between human and animal subjects, assessed during interaction. This was operationalised as a dynamic alignment of physiological signals between individuals during interaction, assessed with any comparative statistical evaluation between variations of human and animal physiological signals.\\
Co-modulation analysis outcomes were classified into three categories: significant, partial (e.g., limited to specific conditions or subsets of data), or absent, based on the authors' interpretation and report of their statistical analyses.\\
No formal assessment of the validity of co-modulation analysis methods was performed; all methods were accepted accordingly to the eligibility criteria of this review.\\
When multiple co-modulation analyses were reported within the same study, the primary outcome reported in the synthesis was defined according to a predefined hierarchy, based on the complexity of the investigated relationship between physiological variables: time-series coupling analyses were prioritized, followed by regression and correlation analyses. \\
When multiple physiological measures analysis results were reported within a study, only the presence of statistically significant co-modulation between human and animal signals for each and every parameter was classified as a "significant" outcome. The presence of statistically significant co-modulation between human and animal signals for a subset of parameters was classified as a "partial" outcome.\\
\subsubsection*{Other Variables}
\label{data items - other variables}
In addition to the primary outcome, the following data items were extracted from each included study: \textit{Name; Year; Animal species involved; Minimum number of human and animal participants; Non healthy (human) / non wild-type (animal) participants; Interaction Context; Behavioural Measures; Measured physiological parameter(s); Sampling; Index used; Analysis description; Data analysis category; Outcome}.\\
When information was missing or unclear, no assumptions were made and the corresponding data fields were left blank. All extracted variables, including missing data, are documented in the data collection form. Definitions and coding rules for each variable are provided in the accompanying data dictionary (see: Supplementary File S3, Supplementary Document S4).

\subsection{Risk of Bias Assessment and Effect Measures}
\label{risk of bias assesssment}
The risk of bias assessment was conducted using a categorization derived and adapted from the ROBINS-I V2 tool (Risk Of Bias In Non-randomized Studies - of Interventions \citep{sterneROBINSIToolAssessing2016}, Version 2 (\url{https://www.riskofbias.info/welcome/robins-i-v2}).\\
Given the heterogeneity of the included studies and the specific context of human-animal physiological and behavioural research, the original structure of the tool was modified to better align with the methodological characteristics of the analysed studies. Specifically, five evaluation domains were defined (\hyperref[Box2]{Box 2}).

\begin{tcolorbox}[colframe=black, title= Box 2: The Five Domains Used in Risk of Bias Evaluation]
\RaggedRight
\footnotesize
\label{Box2}
\textbf{Bias due to confounding:} includes the presence of uncontrolled variables that may influence the association between intervention and outcome.

\textbf{Bias in sample selection:} assesses the representativeness, randomization, and generalizability of the participants.

\textbf{Bias in physiological measurement:} considers the validity, reliability, and standardization of instruments and data collection protocols.

\textbf{Bias in statistical analysis:} evaluates the appropriateness of statistical tests, assumption checking, and use of corrections.

\textbf{Bias in outcome reporting:} assesses transparency in reporting results, including non-significant findings, and consistency with stated objectives.
\end{tcolorbox}
\normalsize

Each domain was rated qualitatively (Low, Moderate, High), with a textual justification. The overall risk of bias was categorized according to a predefined rule: studies with all domains rated as Low were classified as 'Low'; those with at least one Moderate rating as 'Low to Moderate'; those with at least two Moderate ratings as 'Moderate'; those with at least one High rating as 'Moderate to High'; those with and at least two High ratings as 'High'.
This adapted categorization enabled a context-sensitive evaluation of risk of bias, preserving the conceptual structure of ROBINS-I V2 while tailoring it to the specific needs of the review. Discrepancies were resolved through discussion. No automation tools were used in the assessment process.\\
A unified assessment of the risk of bias for the included studies is provided in the \hyperref[characteristics of contributing studies]{"Risk of Bias in Contributing Studies"} section.
The individual assessments for each study are presented in the Supplementary File S5.\\
\label{effect measures}
No standardized effect measures (e.g., risk ratios, mean differences) were used, as the review focused on the categorical outcome (i.e., presence, partial presence or absence of co-modulation), rather than on numerical effect estimates. No thresholds or re-expression of effect measures were thus applied.

\subsection{Synthesis Methods}
\label{synthesis methods - eligibility for synthesis}
\subsubsection*{Eligibility for Synthesis.}  
All included studies were considered eligible for synthesis, regardless of species, context of interaction, or physiological parameter measured. No subgrouping or exclusion was applied at this stage. Studies assessed as being at high risk of bias (see: Supplementary File S5) were included in the synthesis; however, their risk of bias was explicitly acknowledged and considered during the interpretation of the results.

\subsubsection*{Tabulation and Graphical Methods.}
\label{synthesis methods - tabulation and graphical methods}
The results are reported in narrative form in the manuscript and further detailed in supplementary tables.\\
For each synthesis dimension --- data analysis method, interaction context, and measured physiological parameter --- supplementary tables report the distribution of co-modulation outcomes and the corresponding risk of bias assessments.
The supplementary table for the "Data Analysis Method" dimension (see: Supplementary Table S6) is ordered according to a predefined hierarchy, based on the complexity of the investigated relationship between physiological variables.\\
The supplementary table for the "Interaction Context" dimension (see: Supplementary Table S7) includes a further subdivision based on the animal species (dog, horse, others) feature. Additionally, the presence of non-healthy humans or non-wild-type animals is reported for each subcategory.\\
In a similar manner, the supplementary table for the "Physiological Parameter" dimension includes an additional subdivision based on the animal species feature and reports the interaction contexts for each subcategory. This facilitates a more precise comparison across studies (see: Supplementary Table S8). Moreover, within Supplementary Table 8, physiological parameters were categorized and reported, based on the biological domain involved and without implying any hierarchy, as follows:
\small
\begin{itemize}
  \item Cardiac Activity (e.g., heart rate variability (HRV), beats per minute (bpm)
  \item Hormones (e.g., cortisol, oxytocin)
  \item Multiparameter (combinations of cardiac and hormonal measures)
  \item Other (e.g., EEG, breathing rate, Testosterone, CgA, ACTH, b-endorphin, epinephrine, norepinephrine, T3, T4, biochemical parameters (Na,K, CREA, urea, TP, ALB, Mg , AP, CK, ALT, AST), haematological parameters (WBC, RBC, HGB, HCT, MCV, MCH, MCHC, PLT, LYM, GRA)
\end{itemize}
\normalsize
The full names of all abbreviations are listed in the Supplementary Document S4.
Analogously, within Figure \ref{Fig2}, physiological parameters were grouped, based on the sampling modality, as: Cardiac activity,  Molecular Parameters, Cardiac Activity and Molecular Parameters and Others.
To further explore the heterogeneity of the included studies along the "Physiological Parameter" dimension, a supplementary table was compiled to report the occurrence of different sampling methods for molecular physiological parameters (see: Supplementary Table S9). In addition, two grouped bar plots were generated to visualize the number of studies per physiological parameter (non grouped), stratified respectively by interaction context and data analysis method (see: Supplementary Figure S10).\\
\label{synthesis methods - synthesis methods}
All narrative syntheses, structured tables and graphical visualisations were used to present the distribution of outcomes across key methodological dimensions without drawing inferential conclusions, as the performed narrative synthesis does not aim to quantify effect sizes or test hypotheses.

\subsection{Methods to Explore Heterogeneity and Sensitivity Analyses}
\label{heterogeneity and sensitivity analyses}
The descriptive approach of this review was chosen to reflect the heterogeneity of study designs and analytical approaches, and to facilitate comparison across subsets of studies. No formal statistical models or heterogeneity metrics were applied. No sensitivity analyses were conducted, as the synthesis was descriptive and based on categorical outcomes without quantitative pooling.\\
\subsection{Reporting Bias and Certainty Assessment}
\label{reporting bias assessment}
No formal assessment of reporting bias was conducted.
All outcomes were extracted as reported by the original study authors. While statistical significance and effect directions were often reported in the primary studies, no selection or filtering was applied based on these characteristics during data extraction or synthesis. Nonetheless, the risk of bias assessment was conducted using a categorization derived and adapted from the ROBINS-I V2 tool that included an explicit evaluation of the risk of bias in outcome reporting (see:  \hyperref[Box2]{Box 2}, Supplementary File S5).\\
Risk of bias due to missing results has not been assessed due to the fact that each and every entry reported explicitly the outcome of interest (presence, partial presence or absence of physiological co-modulation). No conversion or manipulation of the reported outcomes was performed prior to data extraction.
\label{certainty assessment}
No formal assessment of the certainty (or confidence) in the body of evidence was conducted. 

\section{Results}
\label{Results}

\subsection{Individual Studies Characteristics and Results}
\label{study characteristics}
Characteristics of included studies, as well as a brief summary of each one, are available as Supplementary Material (see: Supplementary Table S11, Supplementary Document S12).
\label{results of individual studies}
Individual study results are presented in a supplementary table, which provides a summary of the measured parameters, analysis methods, and outcomes for each included study (see: Supplementary Table S13). 

\subsection{Risk of Bias in Contributing Studies}
\label{characteristics of contributing studies}
The performed narrative synthesis is based on the complete set of included studies. Accordingly, a unified assessment of risk of bias is provided.
\label{risk of bias in studies} 
Additionally, risk of bias judgments for each domain and study, with detailed justifications for each judgment, are provided as a supplementary material (see: Supplementary Table S5).\\
The most frequent rating for the "Bias due to confounding" domain was "Moderate". This was primarily attributed to the lack of control over individual-level variables such as prior experience with animals, emotional state, and physical fitness. In several studies, the use of a single animal in interaction protocols further limited generalizability and introduced potential confounding effects specific to that individual (see: Supplementary Table S5).\\
The "Bias in sample selection" domain was also most frequently rated as "Moderate". Many studies relied on self-selected participants from homogeneous populations, often with limited demographic diversity. Small sample sizes and the use of convenience sampling were common, and in some cases, the inclusion of single-individual animal models further restricted the external validity of the findings.\\
In contrast, physiological measurement was generally robust across studies, with most receiving a "Low" risk of bias rating. Validated devices and standardized protocols were widely used for the collection of cardiac activity, cortisol and oxytocin. Signal processing and artifact removal procedures were typically well-documented and appropriate.\\
Statistical analysis was similarly strong, with most studies employing suitable models and tests, and explicitly addressing their assumptions.\\
Outcome reporting was generally clear, with most studies presenting both significant and non-significant results and noting limitations. However, the lack of pre-registration raises concerns about selective emphasis on positives. Overall, while internal validity was supported by sound methods, external validity and confounding control remain major limitations.\\
Many studies relied on small and homogeneous samples. Moreover, the animal species involved were almost exclusively domesticated, limiting the applicability of findings to broader human-animal interaction contexts.\\
Interaction protocols varied considerably in duration, setting, and type of engagement, and were often tailored to specific experimental aims rather than standardized across studies. Similarly, control conditions intended to represent “Non-interaction” baselines were highly heterogeneous and context-dependent, reflecting the constraints of individual designs. These variations complicate cross-study comparisons and may introduce uncontrolled variability in physiological responses.\\
These limitations, while relevant, do not invalidate the present synthesis, which aims to explore the presence and nature of experimental evidence for physiological co-modulation in human-animal interactions. Rather than quantifying a specific effect size, the performed narrative synthesis of experimental evidence maps the available literature and evaluates its methodological heterogeneity across studies.\\ 

\subsection{Results of Syntheses}
\label{results of statistical syntheses}

\subsubsection*{Data Analysis Methods Across Studies}
Nine studies performed Time-Series Coupling analysis. Eight of them reported significant co-modulation \citep{guidiWearableSystemEvaluation2016d, lanataQuantitativeHeartbeatCoupling2016g, lanataRoleNonlinearCoupling2017b,  baldwinPhysiologicalBehavioralBenefits2021b, baldwinEffectsEquineInteraction2023b, callaraUnveilingDirectionalPhysiological2024b, renDisruptedHumanDog2024b, wienholdRelationshipEarlyLife2025b}.
One work reported absence of co-modulation \citep{holderExploringDynamicsCanineAssisted2024j}.\\

Eight studies performed Regression analyses. In particular, one study employed Structural Equation Modelling (SEM) and reported significant co-modulation \citep{buttnerEvidenceSynchronizationHormonal2015b}.\\
Three studies employed General Linear Mixed Models (GLMMs); one reported significant co-modulation \citep{nomotoDogsBreathRhythm2024b}, one reported partial co-modulation \citep{gnanadesikanEffectsHumananimalInteraction2024h}, one reported absence of co-modulation \citep{harvieDoesStressRun2021f}.\\
One study employed Generalized Linear Models (GLMs) and reported significant co-modulation \citep{sundmanLongtermStressLevels2019d}.\\
Three studies employed (univariate or multivariate) linear regressions; all of them reported significant co-modulation \citep{jonesInterspeciesHormonalInteractions2006b, byrneEmpathyApathyInvestigating2024b, koskelaBehavioralEmotionalComodulation2024k}.\\

Twenty studies performed correlation analyses. In particular, one study performed a Cross-Correlation analysis and reported significant co-modulation \citep{yorkeEquineassistedTherapyIts2013d}.\\
Three studies performed Time Series Correlation analyses and reported significant co-modulation \citep{katayamaEmotionalContagionHumans2019b, nagasawaDogsShowedLower2023b, mcduffeePsychophysiologicalEffectsEquinefacilitated2024c}.\\
Sixteen studies performed discrete-time correlation analyses. Four of them reported significant co-modulation outcomes \citep{nagasawaOxytocingazePositiveLoop2015a, ryanPhysiologicalIndicatorsAttachment2019d, wojtasSalivaryCortisolInteractions2020b, naberHeartRateSalivary2025b}. \\
Nine of them reported partial co-modulation \citep{handlinAssociationsPsychologicalCharacteristics2012d, janczarekEmotionalReactionsHorses2013b, strzelecSalivaryCortisolLevels2013b, hockenhullExploringSynchronicityHeart2015b, kangInfluenceHorseRider2016b, griggAssessingRelationshipEmotional2022f, wojtasAreHairCortisol2022j, friendPhysiologyHumanhorseInteractions2023e, risvanliEffectVictoryDefeat2025b}.\\
Three of them reported absence of co-modulation \citep{schoberlEffectsOwnerDog2012b, schoberlPsychobiologicalFactorsAffecting2017d, rankinsHeartRateVariability2025d}.\\

The Supplementary Table S6 summarizes results of the narrative synthesis for the "Data Analysis Method" dimension, providing a structured overview of the co-modulation outcomes and corresponding risk of bias assessments across all included studies, grouped by analytical approach.\\
To thoroughly represent the methodological heterogeneity observed, particularly in terms of experimental design and statistical modelling, a brief narrative description of each study is presented in the Supplementary Document S12. These narrative descriptions follow the same order of presentation as the Supplementary Table S6. In addition, studies are listed chronologically by year of publication and sorted by decreasing overall risk of bias to facilitate interpretability and comparison.\\

\subsubsection*{Contexts of Interaction Across Studies}
Twelve studies were conducted in the context of Animal Assisted Intervention (AAI) or Animal Assisted Therapy (AAT), each of them enrolled horses. Ten of these studies reported significant co-modulation outcomes \citep{yorkeEquineassistedTherapyIts2013d,  guidiWearableSystemEvaluation2016d, lanataQuantitativeHeartbeatCoupling2016g, lanataRoleNonlinearCoupling2017b, baldwinPhysiologicalBehavioralBenefits2021b, baldwinEffectsEquineInteraction2023b, mcduffeePsychophysiologicalEffectsEquinefacilitated2024c, callaraUnveilingDirectionalPhysiological2024b, naberHeartRateSalivary2025b,  wienholdRelationshipEarlyLife2025b}.
One reported partial co-modulation \citep{friendPhysiologyHumanhorseInteractions2023e}. One reported absent co-modulation\citep{rankinsHeartRateVariability2025d}. \\

Sixteen studies were conducted in the context of Companionship, all of them enrolled dogs. Seven of these studies reported significant co-modulation \citep{ryanPhysiologicalIndicatorsAttachment2019d,  katayamaEmotionalContagionHumans2019b, nagasawaDogsShowedLower2023b, nomotoDogsBreathRhythm2024b, byrneEmpathyApathyInvestigating2024b, koskelaBehavioralEmotionalComodulation2024k, renDisruptedHumanDog2024b}. Three of them reported partial co-modulation \citep{handlinAssociationsPsychologicalCharacteristics2012d, griggAssessingRelationshipEmotional2022f, gnanadesikanEffectsHumananimalInteraction2024h}. Four of them reported absent co-modulation \citep{schoberlEffectsOwnerDog2012b, schoberlPsychobiologicalFactorsAffecting2017d, harvieDoesStressRun2021f, holderExploringDynamicsCanineAssisted2024j} One study additionally enrolled wolves and reported significant co-modulation \citep{nagasawaOxytocingazePositiveLoop2015a}. One study additionally enrolled cats and reported partial co-modulation \citep{wojtasAreHairCortisol2022j}.\\

Six studies were conducted in the context of Agonistic Sport, three of them enrolled dogs and reported significant co-modulation \citep{ jonesInterspeciesHormonalInteractions2006b, buttnerEvidenceSynchronizationHormonal2015b, sundmanLongtermStressLevels2019d}; three of them enrolled horses and reported partial co-modulation \citep{strzelecSalivaryCortisolLevels2013b, kangInfluenceHorseRider2016b, risvanliEffectVictoryDefeat2025b}.\\

Two studies were conducted in the context of Non-Agonistic Sport, both of them enrolled horses and reported partial co-modulation \citep{janczarekEmotionalReactionsHorses2013b, hockenhullExploringSynchronicityHeart2015b}.\\

One study was conducted in the context of Working Animals training, enrolling Search and Rescue (SAR) dogs, and reported significant co-modulation \citep{wojtasSalivaryCortisolInteractions2020b}.\\

Results of the synthesis for the "Interaction Context" dimension, are additionally presented in the Supplementary Table S7.

\subsubsection*{Physiological Parameters Across Studies}

\begin{figure*}[htbp]
    \centering
    \includegraphics[width=\textwidth]{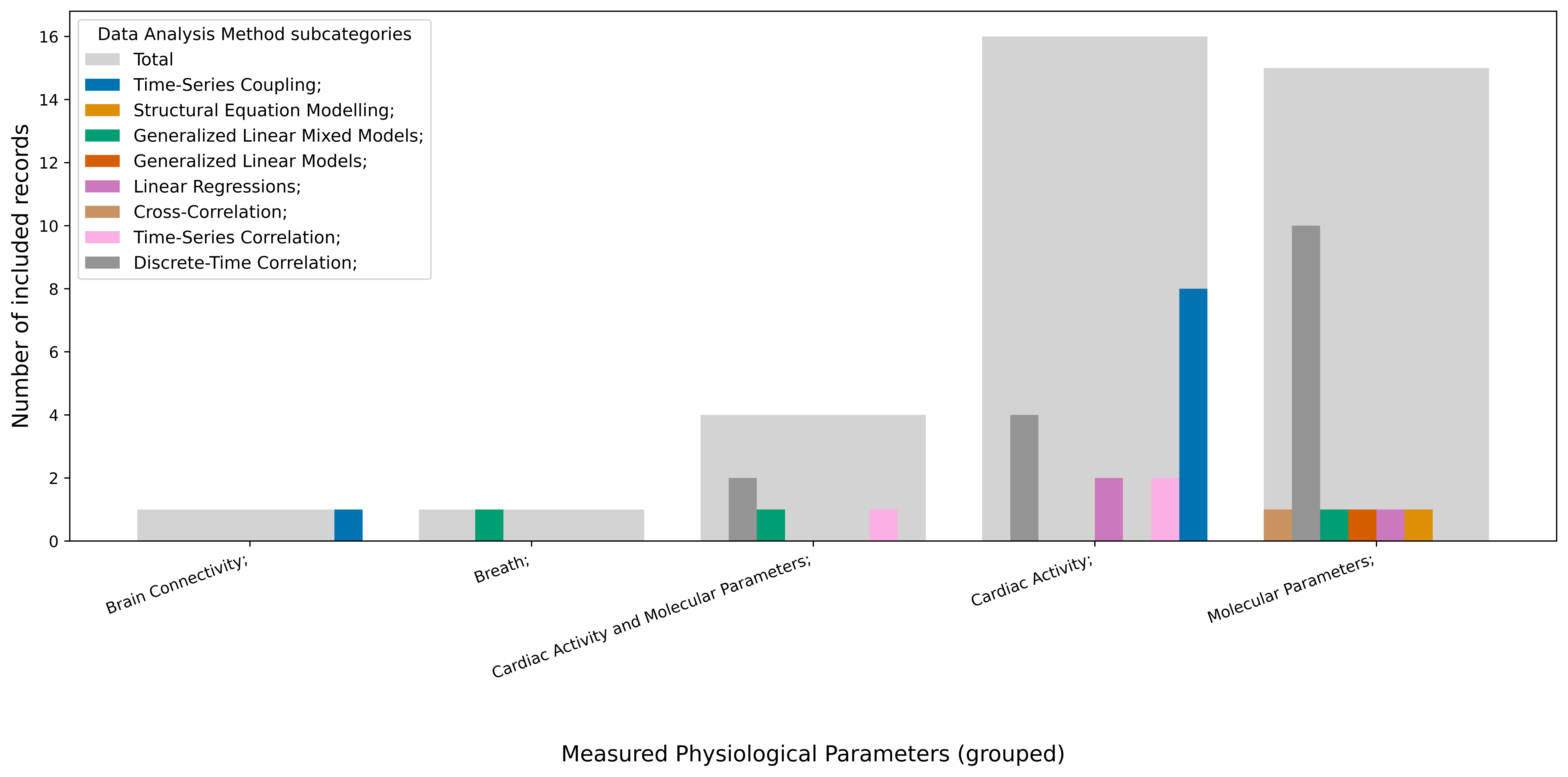}
    
    \vspace{2em}  
    
    \includegraphics[width=\textwidth]{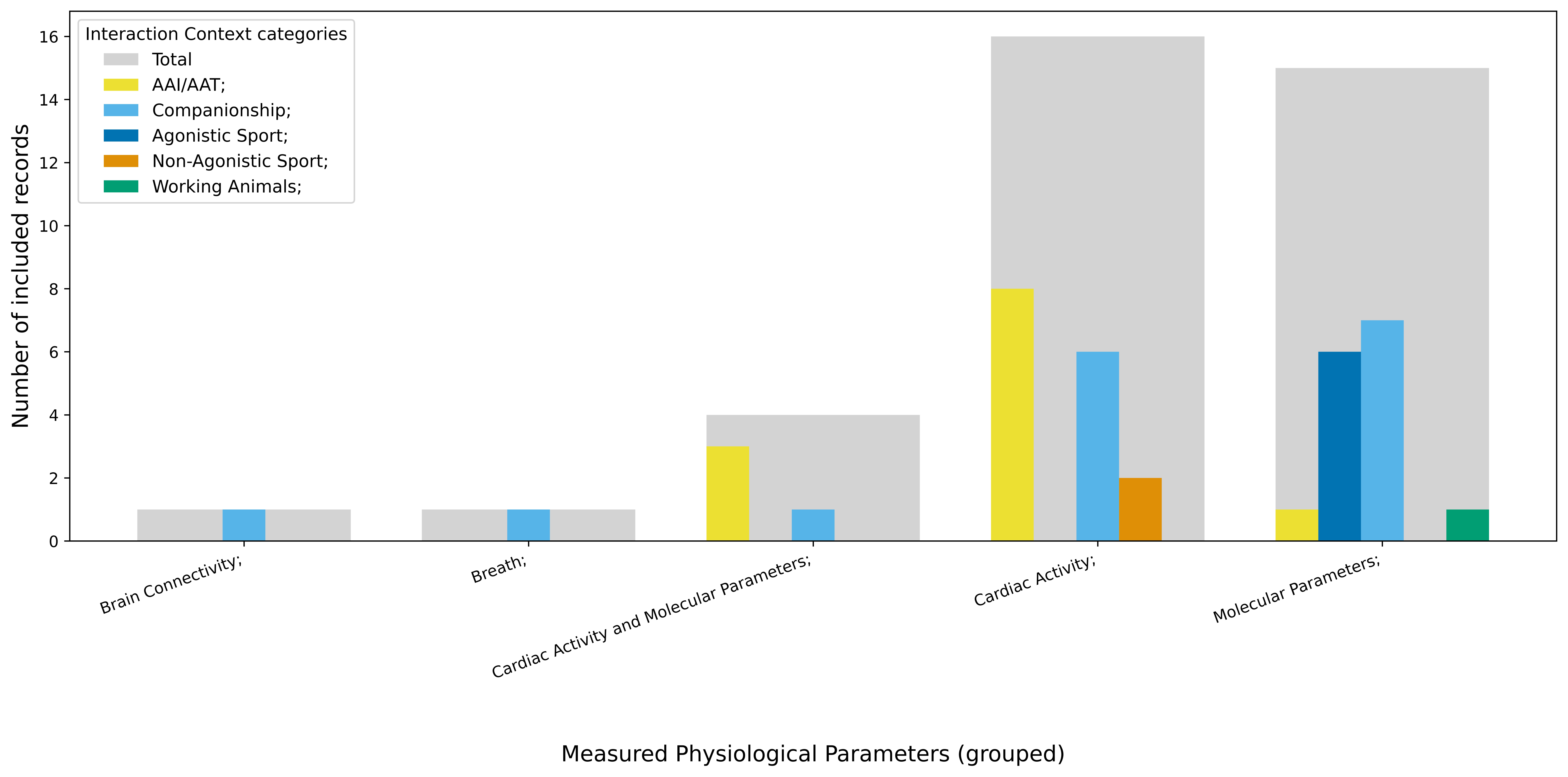}
    \caption{\textbf{Data Analysis Method subcategories and Interaction Context categories across grouped Physiological Parameters} --- Two grouped barplots display the distributions, respectively, of Data Analysis Method subcategories and Interaction Context categories across grouped Physiological Parameters for the retrieved studies. Time-resolved approaches appear more common for continuous signals, while discrete-time methods may be preferred for discrete data. There also seems to be a tendency to use non-invasive techniques in more sensitive contexts, whereas more invasive methods might be employed in less sensitive ones.}
    \label{Fig2}
\end{figure*}

Sixteen studies compared human and animal Cardiac Activity signals. 
Six of them enrolled dogs: four studies reported significant co-modulation \citep{katayamaEmotionalContagionHumans2019b, nagasawaDogsShowedLower2023b, byrneEmpathyApathyInvestigating2024b, koskelaBehavioralEmotionalComodulation2024k}, one reported partial co-modulation \citep{griggAssessingRelationshipEmotional2022f}, one reported absent co-modulation \citep{holderExploringDynamicsCanineAssisted2024j}.\\
Ten of them enrolled horses: seven studies reported significant co-modulation \citep{guidiWearableSystemEvaluation2016d,lanataQuantitativeHeartbeatCoupling2016g,lanataRoleNonlinearCoupling2017b,baldwinPhysiologicalBehavioralBenefits2021b, baldwinEffectsEquineInteraction2023b, callaraUnveilingDirectionalPhysiological2024b,wienholdRelationshipEarlyLife2025b}, two studies reported partial co-modulation \citep{janczarekEmotionalReactionsHorses2013b,hockenhullExploringSynchronicityHeart2015b}, one study reported absent co-modulation \citep{rankinsHeartRateVariability2025d}.\\ Within this group, two studies limited their analysis to heart rate (bpm) characterization, rather than include heart rate variability indices as well. \citep{janczarekEmotionalReactionsHorses2013b,hockenhullExploringSynchronicityHeart2015b}.\\

Eight studies compared human and animal Cortisol Levels. \\
Five of them enrolled dogs: two reported significant co-modulation \citep{sundmanLongtermStressLevels2019d,wojtasSalivaryCortisolInteractions2020b}, two reported absent co-modulation \citep{schoberlEffectsOwnerDog2012b, schoberlPsychobiologicalFactorsAffecting2017d}. The fifth, which additionally enrolled cats, reported partial co-modulation \citep{wojtasAreHairCortisol2022j}.\\
Three of them enrolled horses: one reported significant co-modulation \citep{yorkeEquineassistedTherapyIts2013d}, two reported partial co-modulation \citep{strzelecSalivaryCortisolLevels2013b, kangInfluenceHorseRider2016b}.\\

Two studies compared human and animal Oxytocin Levels, one enrolled dogs and reported partial co-modulation \citep{gnanadesikanEffectsHumananimalInteraction2024h} and the other additionally enrolled wolves and reported significant co-modulation \citep{nagasawaOxytocingazePositiveLoop2015a}.\\

Three studies sampled both Heart Rate and Cortisol levels: one enrolled dogs and reported absent co-modulation \citep{harvieDoesStressRun2021f}; two enrolled horses, one of which reported significant co-modulation \citep{naberHeartRateSalivary2025b}, while the other reported partial co-modulation \citep{friendPhysiologyHumanhorseInteractions2023e}.\\

One study sampled both Cortisol and Oxytocin levels in dogs, reporting partial co-modulation \citep{handlinAssociationsPsychologicalCharacteristics2012d}.\\

One study sampled Heart rate, Cortisol level and Oxytocin level in horses, reporting significant co-modulation \citep{mcduffeePsychophysiologicalEffectsEquinefacilitated2024c}.\\

Two studies measured salivary Testosterone in humans and correlated it to Cortisol levels in dogs, reporting significant co-modulation \citep{jonesInterspeciesHormonalInteractions2006b,buttnerEvidenceSynchronizationHormonal2015b}.\\

One study measured, along with Cortisol levels, salivary CgA levels in both humans and dogs, reporting significant co-modulation \citep{ryanPhysiologicalIndicatorsAttachment2019d}.\\

One study measured, in both humans and horses for each parameter, the plasmatic concentration of many hormones (Cortisol, ACTH, b-endorphin, epinephrine, norepinephrine, T3, T4), various other biochemical (Na,K, CREA, urea, TP, ALB, Mg , AP, CK, ALT, AST) and  haematological (WBC, RBC, HGB, HCT, MCV, MCH, MCHC, PLT, LYM, GRA) parameters (expanded forms of all abbreviations are available in the Supplementary Document S4). This study reported partial co-modulation \citep{risvanliEffectVictoryDefeat2025b}.\\

One study measured breath rhythm in dogs and reported significant co-modulation \citep{nomotoDogsBreathRhythm2024b}.\\

One study measured brain connectivity (EEG) in dogs and reported significant co-modulation \citep{renDisruptedHumanDog2024b}.\\

Results of the narrative synthesis for the "Physiological Parameter" dimension, are additionally presented in the Supplementary Table S8.\\  
Additionally, Figure \ref{Fig2} and Supplementary Figure S10 display the distributions of Data Analysis Method subcategories and Interaction Context categories across physiological parameters. Finally, the Supplementary Table S9 presents the different sampling methods for the molecular physiological parameters across studies.

\subsection{Investigations of Heterogeneity Results}
\label{results of investigations of heterogeneity}
Patterns of heterogeneity were explored through structured categorization and descriptive synthesis, rather than through statistical modelling, deemed inappropriate due to the substantial heterogeneity in study designs, outcome measures, and analytical approaches. Instead, the review focuses on mapping conceptual and methodological patterns across studies.\\
In the presentation of synthesis results, particular attention was given to transparently displaying heterogeneity along the PICO-derived dimensions (\hyperref[Box1]{Box 1}), without engaging in interpretative analysis, using supplementary structured tables to facilitate visual comparison across studies.\\
A narrative and interpretative discussion of these patterns and subgroup structures is provided in Section \ref{limitations of the evidence}, where their implications for understanding physiological co-modulation in human-animal interaction are explored in greater depth.\\

\subsection{Certainty, Sensitivity and Reporting Bias Results}
\label{results of sensitivity analyses}
Neither formal sensitivity analyses nor formal certainty assessments were performed, as the synthesis was qualitative. Methodological robustness was addressed by acknowledging the risk of bias associated with included studies. Additionally, heterogeneity in design and analysis was explicitly displayed in the synthesis, ensuring robustness.
The implications of including studies with varying levels of methodological rigour are discussed in greater depth in Section \ref{limitations of the evidence}, where the influence of bias and study design on the observed pattern of co-modulation outcomes is explicitly acknowledged.\\
\label{reporting biases}
No formal assessment of reporting bias was conducted, as the synthesis was qualitative and not meta-analytic. All results were extracted as reported, without filtering by significance or effect. Potential bias from missing results was mitigated through a comprehensive search strategy, including citation tracking and expert consultation, to maximize evidence completeness.\\

\section{Discussion}
\subsection{General Interpretation of the Results}
\label{general interpretation of the results}
\begin{figure}[ht]
    \centering
    \includegraphics[width=0.5\textwidth]{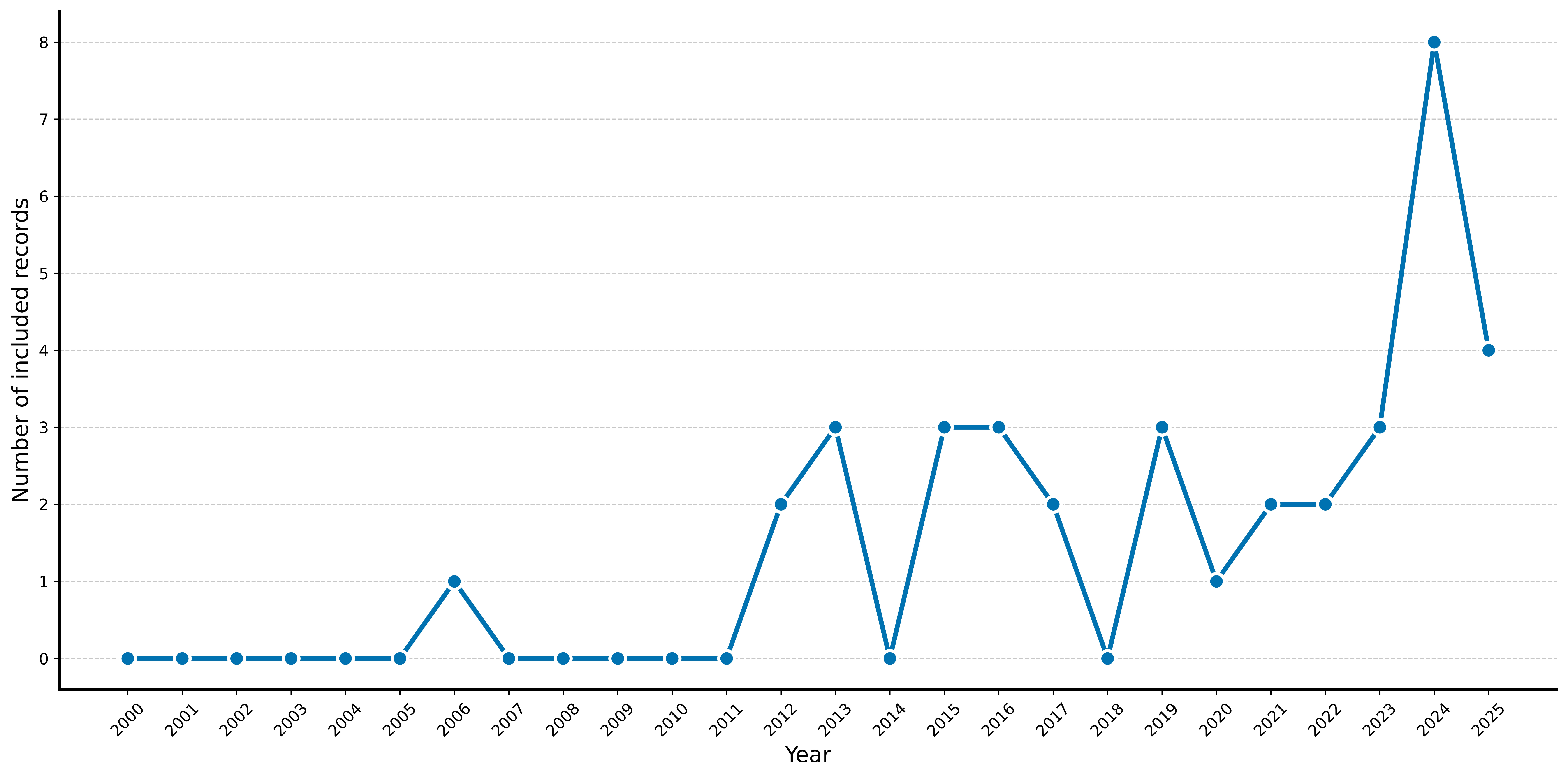}
    \caption{\textbf{Publications per year for the included records} --- The line plot presents the number of publications per year for the records included in this review, highlighting a growth in the number of studies that assess the presence of physiological co-modulation during human-animal interactions.}
    \label{Fig3}
\end{figure}
The narrative synthesis of the categorical co-modulation outcomes for the included studies reveals a growing number of studies (Figure \ref{Fig3}) investigating physiological co-modulation between humans and animals during interaction. Across diverse contexts --- ranging from Animal-Assisted Therapy (AAT) and Companionship to sport and working environments --- most studies report statistically significant associations between human and animal physiological signals, particularly in Cardiac Activity and Hormonal measures (see: Supplementary Table S7, Supplementary Table S8, Supplementary Table S13).\\ 
These findings align with theoretical models of interspecies emotional attunement and mutual regulation, and echo similar patterns observed in human-human dyads, such as parent-child dyads or romantic partners \citep{gollandMereCoPresenceSynchronization2015, kinreichBraintoBrainSynchronyNaturalistic2017, bizzegoStrangersFriendsLovers2019,  gordonPhysiologicalBehavioralSynchrony2020,  linEmotionContagionPhysiological2024}.
However, the evidence remains fragmented. While co-modulation was observed in the majority (22 out of 37) of studies, it was absent (5 out of 37) or only partially present (10 out of 37) in others, especially in studies involving hormonal parameters or less structured interaction settings (see: Supplementary Table S7, Supplementary Table S8).\\ This variability suggests that co-modulation may be context-dependent, influenced by factors such as the type of interaction, the species involved, and the physiological parameter measured. For instance, time-series coupling methods applied to cardiac activity data yielded relatively more consistent evidence of co-modulation than Discrete-Time correlation analyses of hormonal signals (see: Supplementary Table S6, Supplementary Table S8).

\subsection{Limitations of the Evidence}
\label{limitations of the evidence}
Nevertheless, it should be noted that these findings emerge from a highly heterogeneous set of studies which differ substantially in terms of interaction context, animal species, physiological parameters measured, and analytical methods employed (see: Supplementary Table S11). 
This diversity calls for caution in interpretation: the presence of co-modulation cannot yet be considered a generalizable phenomenon across human-animal interactions. In particular, given the lack of consistent evidence for the existence of co-modulation as a general effect, it was neither feasible nor meaningful to quantitatively analyse the distribution of outcomes across synthesis categories in search of trends.\\
Notably, the studies that reported the most consistent evidence of co-modulation —-- particularly those employing time-series coupling techniques —-- were also among those with the highest overall risk of bias (see: Supplementary File S5, Supplementary Table S6), especially in the domains of confounding (e.g., uncontrolled emotional or experiential variables) and sample selection (e.g., small or homogeneous samples). Despite their analytical sophistication, these studies often lacked methodological safeguards that would support generalizability.\\
More broadly, the methodological robustness and risk of bias varied considerably across the included studies (see: Supplementary File S5). 
As detailed in Section \ref{risk of bias in studies}, although physiological measurement and statistical analysis were generally sound, the overall methodological robustness varied, with most studies rated as "Moderate" in the overall risk of bias, and a substantial number rated as "Moderate to High".\\
However, the synthesis suggests several ways in which researchers should approach this evidence prudently, by critically evaluating the methodological assumptions, contextual constraints, and analytical frameworks that shape each study's conclusions.\\

As an example, the terminology used to describe interspecies physiological alignment is often inconsistent. The term “synchrony” is frequently employed in the retrieved papers, yet its operationalization varies widely—from strict time-resolved coupling to broader statistical associations. This ambiguity complicates cross-study comparisons and may obscure the underlying mechanisms of interaction. As proposed by \citep{koskelaBehavioralEmotionalComodulation2024k}, the term “co-modulation”, adopted in this review, may offer a more accurate and inclusive descriptor, capturing mutual physiological influence over extended timescales without implying the use of specific signal processing techniques typically associated with the concept of synchrony.\\

Moreover, as reflected qualitatively in the synthesis results (Figure \ref{Fig2}, Supplementary Figure S10) the specific physiological parameter measured, as well as its sampling modality, strongly influences the feasibility of different co-modulation analysis protocols. Continuous signals such as cardiac activity or EEG are well-suited for time-series coupling techniques, enabling fine-grained analysis of temporal dynamics. In contrast, discrete measures like hormonal (e.g., cortisol, oxytocin) concentrations pose challenges due to their considerably lower temporal resolution. Even within hormonal measures, the biological matrix used 
(saliva, plasma, urine, hair --- see: Supplementary Table S9) affects both the temporal window captured and the biological interpretation of the signal. For example, hair cortisol reflects long-term accumulation, while salivary cortisol captures acute stress responses. In general, the onset and temporal dynamics of the physiological signal vary considerably depending on the sampling modality \citep{schillingReviewNonInvasiveSampling2022} introducing important differences in how co-modulation can be biologically interpreted and generalized across studies.\\ These discrepancies must be taken into account not only when selecting analytical approaches, but especially when attempting to generalize the presence or absence of co-modulation across human-animal interactions and provide a corresponding biological interpretation. Without accounting for the temporal and biological specificity of each parameter, conclusions about the existence of co-modulation risk being context-bound and non-transferable.\\

Additionally, the synthesis results (Figure \ref{Fig2}, Supplementary Figure S10, Supplementary Table S8) qualitatively suggest that the sampling modality may be shaped by the interaction context. For example, in therapeutic settings such as Animal-Assisted Therapy (AAT) or Intervention (AAI), invasive procedures like blood draws may induce stress and compromise the ecological validity of the interaction \citep{suba-bokodiUnconventionalAnimalSpecies2024}. This can alter the physiological signal itself, introducing stress responses unrelated to the interaction, and thereby compromise the validity of co-modulation analysis, which may end up capturing the effects of the sampling procedure rather than interaction dynamics.\\
Conversely, in sport contexts, where blood sampling is already part of routine practice \citep{moreiraDopingDetectionAnimals2021}, plasmatic samples can be obtained without further altering the interaction context. Notably, the establishment of robust connections between human-animal interaction research and agonistic sport environments has the potential to facilitate the creation of scientific datasets of relevance.\\ 
Furthermore, the interaction context influences not only the sampling strategy but also the type of subjects involved (see: Supplementary Table S7). In AAI, human participants are often Non-healthy individuals (e.g., patients with Post Traumatic Stress Disorder (PTSD), physical and/or intellectual disabilities, or age-related conditions), whereas sport studies typically involve healthy, trained individuals.\\
Similarly, the animal species enrolled are predominantly domesticated, primarily dogs and horses. Their domestication has made them central to therapeutic, recreational, and working contexts, which in turn facilitates funding and data availability. Additionally, domesticated animals do offer practical advantages for data collection, particularly in terms of compliance with wearable devices and tolerance to experimental procedures \citep{suba-bokodiUnconventionalAnimalSpecies2024}. Nonetheless, the ecological generalizability of findings remains limited, and future research should aim to include wild and laboratory animals to broaden the scope of inference.\\

Finally, to advance the field, it is essential to design protocols that explicitly test whether significant co-modulation occurs, rather than assuming its presence a priori. In many studies, the main research question is framed around highly specific scenarios, without addressing a broader and more foundational issue: whether significant co-modulation between humans and animals is occurring at all. A more rigorous approach would involve framing the presence of co-modulation as a primary research question, supported by appropriate control conditions and statistical testing.\\
Moreover, the concept of co-modulation should be systematically contextualized with respect to the type of interaction, the experimental setting, the physiological parameters measured, the species involved, and the analytical methods used. This is particularly necessary due to the fact that the development of this field of study seems to be characterised by the adoption of increasingly complex physiological parameters combinations and data analysis procedures (see: Supplementary Figure S14).\\ While this framework, illustrated by the present synthesis, offers one possible way to structure the evidence, it is not intended as exhaustive or prescriptive. Its further development may nonetheless facilitate future generalizations and meta-analyses.\\

\subsection{Limitations of the Review Processes}
\label{limitations of the review processes}
The first key limitation of this review lies in the initial search strategy, which relied on the compound keyword "human animal interaction". This choice was influenced by an implicit availability bias: the expectation that most relevant studies would be situated within the domain of Animal-Assisted Therapy (AAT) or Intervention (AAI), where the term and its acronym (HAI) are commonly used. However, this assumption may have inadvertently reduced the sensitivity of the search, potentially overlooking studies from other important contexts such as Sport and Companionship related ones. The subsequent broadening of the search strategy to include multiple Boolean keyword combinations targeting specific physiological parameters (see: Supplementary Document S1), motivated by the need to enhance sensitivity, may have incidentally contributed to balancing the contextual limitations introduced by the initial keyword choice. Additionally, the use of the Consensus App and of citation tracking helped identify studies not indexed under standard HAI terminology, thereby enhancing the comprehensiveness of the evidence base.\\
Furthermore, as was discussed in the preceding sections (see: \hyperref[Methods]{Methods}, \hyperref[Results]{Results}), it is imperative to acknowledge many additional limitations: the review does not include a quantitative meta-analysis; the outcomes were categorized qualitatively; no pooled effect estimates were calculated; statistical assessments of heterogeneity, as well as sensitivity analyses, and formal evaluations of reporting bias, which are typically employed in meta-analytic frameworks, have not been performed.\\
Additionally, the adopted synthesis dimensions, despite being specifically selected to represent the heterogeneity observed across the PICO framework elements (\hyperref[Box1]{Box 1}), were not pre-specified.\\
Finally, the risk of bias was assessed using a customized framework adapted from ROBINS-I V2, tailored to the specific methodological features of human-animal interaction studies. While this approach aimed to preserve conceptual rigour, it does not follow a standardized or validated protocol.\\

\subsection{Implications of the Results}
\label{implications of the results}
In conclusion, this review presents preliminary evidence of physiological co-modulation between humans and animals, without determining whether the phenomenon depends on specific combinations of interaction context, species, physiological parameter, and analytical method. Both interpretations remain plausible: co-modulation may be a general feature of interspecies interaction, or it may emerge only under certain biological and methodological conditions.\\
While the synthesis supports the plausibility of physiological co-modulation, its application in real-world settings should remain cautious. In therapeutic contexts, for example, it is premature to use co-modulation as a parameter for session tuning or outcome evaluation. Similarly, in Companionship studies, interpreting physiological alignment as a proxy for bonding still lacks definitive empirical grounding. Until the phenomenon is better characterised, it is advisable for practical implementations to avoid assuming a priori its occurrence, as well as treating it as an universally established mechanism.\\
Future studies, rather than assuming generalizability, should aim to explicitly test the presence of co-modulation as a primary outcome, using appropriate control conditions and analytical methods.\\
Expanding the scope to include Non-domesticated species, Non-therapeutic contexts, and datasets structured from sport environments may help clarify the conditions under which co-modulation occurs. \\
At the policy level, the current evidence base remains insufficient to support standardized frameworks for physiological monitoring in human-animal interaction. Further conceptual and methodological consolidation is needed before translating these findings into practice or regulation.\\

\section{Other Information}
\subsection{Registration and Protocol}
\label{Registration and protocol}
This systematic review was not formally registered in a public repository. The review was conceived as an exploratory synthesis of emerging literature in a novel research area. Given its preliminary nature and evolving scope, formal registration was not pursued.
However, all methodological decisions were transparently documented, in accordance with PRISMA 2020 guidelines.
\subsection{Support}
\label{support}
We acknowledge financial support under the National Recovery and Resilience Plan (NRRP), Mission 4, Component 2, Investment 1.1, Call for tender No. 104 published on 02/02/2022 by the Italian Ministry of University and Research (MUR), funded by the European Union (NextGenerationEU). Project Title: "One welfare, one emotion: a look Inside the interaction. Bioengineering solution for the human-horse emotional transfer in neuro-psychological, and social perspectives. (OneFeeL)"; CUP: B53C24007430006; Grant Assignment Decree No. 104 adopted on 02/02/2022 by the Italian Ministry of Ministry of University and Research (MUR).
\subsection{Competing Interests}
\label{Competing interests}
No competing interests are reported. There are no personal, professional, or financial relationships that could be perceived as influencing the conduct or reporting of this systematic review.
\subsection{Declaration of generative AI and AI-assisted technologies in the writing process}
During the preparation of this manuscript, the authors employed Copilot (\url{https://copilot.microsoft.com/}, desktop version) to assist in locating specific excerpts within the full texts of the included studies during data collection, the actual presence of the retrieved citations was manually checked.\\
Moreover, the DeepL (\url{https://www.deepl.com/it/translator}) desktop tool was utilised to enhance the clarity and fluency of the English composition of this review.
All content generated with these tools was subsequently manually reviewed and edited by the authors, who assume full responsibility for the final version of the manuscript.
\subsection{Availability of Data, Code and Other Materials}
\label{Availability of data, code and other materials}
The following materials are provided in the supplementary section:\\
The PRISMA search strategy document (S1);\\ The Zotero collection of screened records, organized by source and screening stage (S2);\\ The structured data collection form, including the extraction template and all data used for synthesis (S3);\\ The accompanying data dictionary defining all variables and coding rules (S4);\\ The risk of bias assessment table for all included studies (S5);\\ The Data Analysis synthesis results structured table (S6);\\ The Interaction Context synthesis results structured table (S7);\\ The Physiological Parameters synthesis results structured table (S8);\\ The molecular parameters sampling categories table (S9);\\ An additional figure, analogous to Figure \ref{Fig2}, with non-grouped physiological parameters (S10);\\ The table summarizing characteristics of individual studies (S11);\\ The narrative summary of each study (S12);\\ The table summarizing results of individual studies (S13);\\ A second additional figure, similar to Figure \ref{Fig3}, displaying publications per year for different Physiological Parameters groups and Data Analysis categories (S14);\\ The PRISMA checklist for this review (S15);\\ The PRISMA abstract checklist for this review (S16);\\ The PRISMA search strategy document checklist for this review (S17).

\footnotesize
\bibliography{Bibliography}

\end{document}